\documentclass[10pt,aps,twocolumn,secnumarabic,balancelastpage,amsmath,amssymb,nofootinbib,hyperref=pdftex,prx,citeautoscript,showpacs]{revtex4-1}

\usepackage{graphics}
\usepackage[pdftex]{graphicx}
\usepackage{tabularx}
\graphicspath{{./Figures/}}
\usepackage{longtable}
\usepackage{amsmath}
\usepackage{color,soul}
\usepackage{braket}
\usepackage{float}

\makeatletter
\def\p@subsection{\thesection\,}
\makeatother

\usepackage{hyperref}
\hypersetup{colorlinks=true,linkcolor=blue,anchorcolor=blue,citecolor=blue,filecolor=blue,urlcolor=blue,bookmarksnumbered=true,pdfview=FitB}

\begin{document}

\title{Metallization of Rashba wire by superconducting layer in the strong-proximity regime}
\author{Christopher Reeg}
\author{Daniel Loss}
\author{Jelena Klinovaja}
\affiliation{Department of Physics, University of Basel, Klingelbergstrasse 82, CH-4056 Basel, Switzerland}
\date{\today}
\begin{abstract}
Semiconducting quantum wires  defined within two-dimensional electron gases and strongly coupled to thin superconducting layers  have been extensively explored in  recent experiments as promising platforms to host Majorana bound states. 
We study numerically such a geometry, consisting of a quasi-one-dimensional wire coupled to a disordered three-dimensional superconducting layer. We find that, in the strong coupling limit of a sizable proximity-induced superconducting gap, all transverse subbands of the wire are significantly shifted in energy relative to the chemical potential of the wire. For the lowest subband, this band shift is comparable in magnitude to the spacing between quantized levels that arise due to the finite thickness of the superconductor (which typically is $\sim500$ meV for a 10-nm-thick layer of Aluminum); in higher subbands, the band shift is much larger. Additionally, we show that the width of the system, which is usually much larger than the thickness, and moderate disorder within the superconductor have almost no impact on the induced gap or band shift. We provide a detailed discussion of the ramifications of our results, arguing that a huge band shift and significant renormalization of semiconducting material parameters in the strong-coupling limit make it challenging to realize a topological phase in such a setup, as the strong coupling to the superconductor essentially metallizes the semiconductor. This metallization of the semiconductor can be tested experimentally through the measurement of the band shift.

\end{abstract}

\maketitle

\section{Introduction} \label{secIntro}

The search for Majorana fermions in various condensed matter systems has intensified considerably in recent years \cite{Alicea:2012}. Among the most promising proposals for realizing these exotic states involve coupling a conventional superconductor to a topological insulator \cite{Fu:2008,Fu:2009,Hasan:2010,Qi:2011,Liu:2011,Wiedenmann:2016}, an atomic magnetic chain \cite{NadjPerge:2014,Ruby:2015,Pawlak:2016,Klinovaja:2013,Vazifeh:2013,Braunecker:2013,NadjPerge:2013,Pientka:2013,Awoga:2017}, or a semiconductor with strong spin-orbit coupling \cite{Sato:2009,Sato:2009b,Lutchyn:2010,Oreg:2010,Sau:2010,Alicea:2010,Chevallier:2012,Halperin:2012,Sticlet:2012,Prada:2012,Dominguez:2012,Klinovaja:2012,Nakosai:2013,DeGottardi:2013,Weithofer:2013,Weithofer:2014,Vernek:2014,Maier:2014,Thakurathi:2015,Dmytruk:2015,Dominguez:2017,Maska:2017}. The first generation of experiments on semiconductor/superconductor hybrid structures, which showed zero-bias peaks in the tunneling conductance of nanowires, were plagued by significant subgap conductance \cite{Mourik:2012,Deng:2012,Das:2012,Churchill:2013,Finck:2013,Takei:2013,Stanescu:2014}. This led to the development of epitaxially grown thin shells of superconducting Aluminum (Al) that form a very strong and uniform contact with InAs or InSb nanowires, thus ensuring robust proximity couplings and hard induced superconducting gaps that are nearly as large as the gap of the Al layer \cite{Chang:2015,Albrecht:2016,Deng:2016,Gazibegovic:2017,Zhang:2017_2,Vaitiekenas:2017,Deng:2017}. The epitaxial growth of Al has also been extended to InAs two-dimensional electron gases (2DEGs), with the hope that such systems can be used to form complex networks of Majorana fermions \cite{Kjaergaard:2016,Shabani:2016,Kjaergaard:2017,Suominen:2017,Nichele:2017}.

With the experiments shifting to the strong-coupling regime, the proximity effect in topological setups has gained renewed attention. It is well known that as the coupling between the superconductor and semiconductor is enhanced, the electron wave function acquires a larger weight within the superconductor, thus leading to a larger proximity-induced gap ($E_g$) and a renormalization of semiconducting material parameters such as $g$-factor, spin-orbit splitting ($E_{so}$), and effective mass ($m^*$) \cite{Sau:2010prox,Stanescu:2010,Potter:2011,Tkachov:2013,Zyuzin:2013,Cole:2015,vanHeck:2016,Hell:2017,Stanescu:2017,Liu:2017,Reeg:2017_2}. This result can be obtained analytically in the limit of a single 1D or 2D semiconducting subband coupled to a clean 3D bulk superconductor by ``integrating out" the superconducting degrees of freedom to obtain an effective self-energy describing induced superconductivity. In this description, the relevant superconducting energy scale determining the strength of the proximity effect is the gap $\Delta$ \cite{strongcoupling}, and all physics of the proximity effect should occur on this small energy scale. Despite such theories being employed frequently to describe the experiments utilizing thin epitaxial superconducting layers \cite{Cole:2015,vanHeck:2016,Deng:2016,Zhang:2017_2,Stanescu:2017,Liu:2017,Hell:2017}, their applicability is unclear because the experimental setup consists of multiple wire channels and a thin disordered superconducting layer (rather than a clean bulk  superconductor).

\begin{figure}[t!]
\centering
\includegraphics[width=\linewidth]{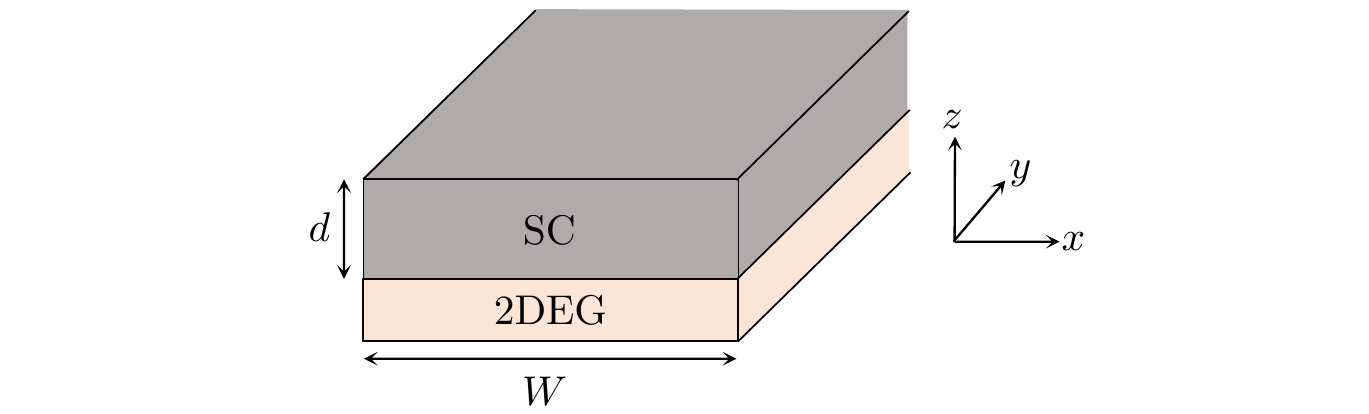}
\caption{\label{experiment} A quantum wire is lithographically defined within a semiconducting 2DEG and coupled to a superconducting layer with width $W$ and thickness $d$, as studied in Refs.~\cite{Suominen:2017,Nichele:2017}.
}
\end{figure}

The finite thickness ($d$) of the superconducting layer was incorporated analytically in Ref.~\cite{Reeg:2017_3} in an attempt to better describe the experimental setup. This finite thickness introduces a large energy scale given by the level spacing $\pi \hbar v_{Fs}/d$ ($v_{Fs}$ is the Fermi velocity of the superconductor), and it was shown that the relevant superconducting energy scale determining the strength of the proximity effect in this case is the level spacing rather than the gap. Thus, in the thin-layer limit, $\hbar v_{Fs}/d\gg\Delta$, a much stronger proximity coupling is required in order to open a gap in the wire. Most importantly, in addition to the usual parameter renormalization, the strong-coupling limit is also accompanied by a large  band shift in the semiconducting wire that is comparable to the level spacing $\pi \hbar v_{Fs}/d$ \cite{Reeg:2017_3}. In stark contrast, such a band shift is completely absent in the  case of a bulk superconductor \cite{Sau:2010prox,Stanescu:2010,Potter:2011,Tkachov:2013,Zyuzin:2013,Cole:2015,vanHeck:2016,Hell:2017,Stanescu:2017,Liu:2017,Reeg:2017_2}. 

However, Ref.~\cite{Reeg:2017_3} studied an idealized model of a single 1D semiconducting subband coupled to a clean 2D superconductor. The conclusion that a very strong proximity coupling is required to open a sizable gap is strongly dependent on the assumption of momentum conservation within the tunneling process. In a realistic experimental setup with superconductor thickness $d\sim10$ nm and width $W\sim50$ nm \cite{Suominen:2017,Nichele:2017} (see Fig.~{\ref{experiment}}), there are $\sim10^4$ occupied superconducting subbands; in the presence of unavoidable disorder within the superconducting layer, which breaks translational symmetry and therefore removes momentum conservation, one could expect that subbands of the wire can much more easily couple to the many superconducting subbands that lie at the Fermi level, thus significantly reducing the coupling strength required to open a sizable gap in the wire.
Additionally, if there is a large band shift that exceeds the transverse level spacing of the wire, then higher subbands become increasingly important and a single-band analytical model like that studied in Ref.~\cite{Reeg:2017_3} is insufficient. While there have been several works to investigate the stability of topological superconducting phases in multiband wires \cite{Wimmer:2010,Stanescu:2011,Potter:2010,Potter:2011b,Lutchyn:2011,Lutchyn:2011b,Zhou:2011,Law:2011,Kells:2012,Pientka:2012,Gibertini:2012,Rainis:2013,Rieder:2014,Pekerten:2017}, there are no systematic studies of the  proximity effect in the strong-coupling limit. It is the goal of this work to provide such a study.

In this paper, we numerically study the proximity effect in a quasi-1D quantum wire that is defined within a semiconducting 2DEG and strongly coupled to a thin disordered superconducting layer with thickness $d$ and width $W$, as shown in Fig.~\ref{experiment}. First, we show that in the strong-coupling limit, which is characterized by the wire having a proximity-induced gap $E_g$ that is comparable to the gap $\Delta$ of the superconductor, all transverse subbands of the wire are significantly shifted with respect to their positions in the absence of coupling, and all semiconducting material parameters (such as effective mass $m^*$, spin-orbit splitting $E_{so}$, and $g$-factor) are significantly renormalized toward their values in the superconductor. Next, we study in detail the role played by both the finite width $W$ of the system and disorder within the superconductor. We find, quite surprisingly, that neither the finite width $W$ nor moderate levels of disorder have a substantial impact on the proximity effect. For the specific case of an InAs quantum wire coupled to a thin epitaxial layer of Al, we show that the semiconducting wire becomes metallized by the superconductor. This metallization is characterized by the occupation of many transverse subbands in the wire (thus pushing the system far from the desired 1D limit) and a significant renormalization of the semiconducting material parameters. Additionally, we discuss the challenges involved in realizing a topological phase in the metallized limit, arguing that the ability to do so is unlikely but is also largely device dependent, and propose how to experimentally test our theory. Our results presented in this work suggest that it is more promising to search for Majorana fermions in systems with a weak proximity coupling (such as nanowires sputtered by a thick superconducting slab \cite{Mourik:2012}) than in systems with  strong proximity coupling accompanied by a large band shift.

%

While our focus is directed primarily toward engineering topological superconductivity in 1D systems, 
the study of a strong proximity coupling between a semiconductor with Rashba spin-orbit interaction (SOI) and an $s$-wave superconductor has far-reaching consequences. In particular, we expect that our results can be extended to studying the proximity effect in topological insulator surface states \cite{Fu:2008,Fu:2009,Hasan:2010,Qi:2011,Liu:2011,Crepin:2015,Dayton:2016,Wiedenmann:2016,Charpentier:2017,Cayao:2017}, odd-frequency triplet superconductivity \cite{BlackSchaffer:2012,Asano:2013,Bergeret:2013,Bergeret:2014,Reeg:2015,Hart:2016,Yang:2017} and magnetoelectric effects \cite{Edelstein:2003,Bobkova:2017} induced by SOI, superconducting spintronics \cite{Buzdin:2005,Bergeret:2005,Eschrig:2011,Linder:2015}, Cooper pair splitting \cite{Byers:1995,Choi:2000,Deutscher:2000,Lesovik:2001,Recher:2001,Yeyati:2007,Shekhter:2016}, as well as various aspects of the superconductor-insulator transition \cite{Bottcher:2017}.

The remainder of the paper is organized as follows. In Sec.~\ref{secModel}, we describe our numerical tight-binding simulation. The results of our calculation for a disordered 3D system are presented in Sec.~\ref{secNumerics}, where we justify that such a system can be realistically described by a clean 2D model. In Sec.~\ref{secEstimates}, we provide a numerical calculation specific to epitaxial Al/InAs experimental setups. We also argue that the metallization of InAs inhibits the ability to observe a 1D topological phase in such a setup and discuss how to experimentally test our theory. Our conclusions are summarized in Sec.~\ref{secConclusion}.

\section{Model} \label{secModel}
We consider the geometry sketched in Fig.~\ref{experiment}, which consists of a quasi-1D semiconducting quantum wire of width $W$ with Rashba SOI, assumed to be lithographically defined within a 2DEG similarly to the devices studied in Refs.~\cite{Suominen:2017,Nichele:2017}, tunnel coupled to a superconducting layer of width $W$ and thickness $d$. We do not consider explicitly the finite thickness of the 2DEG, as we assume that the subband spacing arising from the finite thickness is very large (for the experimental thickness $\sim5$ nm \cite{Shabani:2016}, this is a valid assumption). We describe this setup by a tight-binding Hamiltonian, assuming for now that the system is clean and translationally invariant along its length. The total tight-binding Hamiltonian is given by $H=\sum_kH_{k}$, where $k$ is a conserved momentum along the wire axis; for a given momentum, we consider
\begin{equation} \label{Htot}
H_k=H_k^w+H_k^s+H_k^t.
\end{equation}

The Hamiltonian of the wire is given by
\begin{equation}
\begin{aligned}
H_k^w&=\sum_{x=1}^{W/a}\biggl[b_{x,k}^\dagger\{\xi_{k}^w+(\alpha/a)\sin(ka)\sigma_x-\Delta_Z\sigma_y\}b_{x,k} \\
	&-\{b_{x,k}^\dagger(t_w-i\alpha\sigma_y/2a) b_{x+1,k}+H.c.\}\biggr],
\end{aligned}
\end{equation}
where $b_{x,k}=(b_{x,k,\uparrow},b_{x,k,\downarrow})^T$ is a spinor, $b_{x,k,\sigma}$ annihilates a state of momentum $k$ and spin $\sigma$ at position $x$ within the wire, $t_w$ is the hopping matrix element, and $a$ is the lattice constant. In addition, the Hamiltonian contains a Rashba SOI term \cite{Bychkov:1984,rashba} characterized by the SOI constant $\alpha$ as well as a Zeeman term $\Delta_Z=|g|\mu_BB/2$ caused by an external magnetic field of strength $B$ applied along the wire axis ($g$ is the $g$-factor of the wire and $\mu_B$ is the Bohr magneton). The SOI term induces a spin-orbit splitting $E_{so}$ on each transverse subband of the wire, which is defined as the difference in energy between the crossing point of spin-split bands at $k=0$ and the bottom of the band. Due to the finite width $W$, the spin-orbit splitting is different for each transverse subband; in the limit $W/a=1$, the splitting is given by the usual expression $E_{so}=m^*\alpha^2/2\hbar^2=\alpha^2/4t_wa^2$ \cite{Bychkov:1984,Rainis:2013}, where $m^*$ is the effective mass of the wire (in terms of tight-binding parameters, $m^*=\hbar^2/2t_wa^2$). We take $\xi_k^w=2t_w\{1-\cos(ka)-(1+\alpha^2/8t_w^2a^2)\cos[\pi W/(W+1)]\}-\mu_w$, such that the chemical potential of the wire $\mu_w$ is measured from the Rashba crossing point (at $k=0$) of the lowest transverse subband. 

The Hamiltonian of the superconducting layer is
\begin{align}
H_k^s&=\sum_{x=1}^{W/a}\sum_{z=1}^{d/a}\biggl[c^\dagger_{x,z,k}(\xi_{k}^s-\Delta_Z^s\sigma_y)c_{x,z,k}-\{t_sc_{x,z,k}^\dagger c_{x+1,z,k} \nonumber\\
	&+t_sc_{x,z,k}^\dagger c_{x,z+1,k}+\Delta c_{x,z,-k,\downarrow}^\dagger c_{x,z,k,\uparrow}^\dagger+H.c.\}\biggr],
\end{align}
where $c_{x,z,k,\sigma}$ annihilates a state of momentum $k$ and spin $\sigma$ at position $(x,z)$ within the superconductor, $t_s$ is the hopping matrix element, and $\Delta$ is the pairing potential. The external magnetic field is incorporated in the superconductor through the Zeeman term $\Delta_Z^s=(2/|g|)\Delta_Z$, and we take $\xi_k^s=2t_s\{2-\cos[\pi W/(W+1)]-\cos(ka)\}-\mu_s$, such that the chemical potential of the superconductor $\mu_s$ is measured from the bottom of the lowest subband. 

Finally, tunneling between the wire and superconductor, which is assumed to preserve spin and momentum, is described by
\begin{equation} \label{Ht}
H_k^t=-t\sum_{x=1}^{W/a}\{c_{x,1,k}^\dagger b_{x,k}+H.c.\},
\end{equation}
where $t$ is real and denotes the tunneling strength.

Our model assumes that the chemical potential in the system is fixed externally, and hence any change in particle number can be attributed to the attached leads. However, as any change in particle number that may occur is small compared to the total number of particles in the system (see Sec.~\ref{secTopological}), we expect a negligible deviation from what would be obtained in the case of fixed particle number.

\section{Numerical results} \label{secNumerics}
In this section, we present results obtained numerically. For now, we do not attempt to explicitly model any existing experimental setup; due to the very short Fermi wavelength of the metal, doing so would be extremely expensive computationally. Rather, we focus on deducing various numerical trends that arise when keeping the physical separation of energy scales intact (e.g. $\mu_s\gg\mu_w$). As we will see, these results will allow us to make more quantitative predictions about the experimental setup in the following section.

\subsection{Strong coupling limit} \label{secStrongCoupling}
\begin{figure}[t!]
\centering
\includegraphics[width=\linewidth]{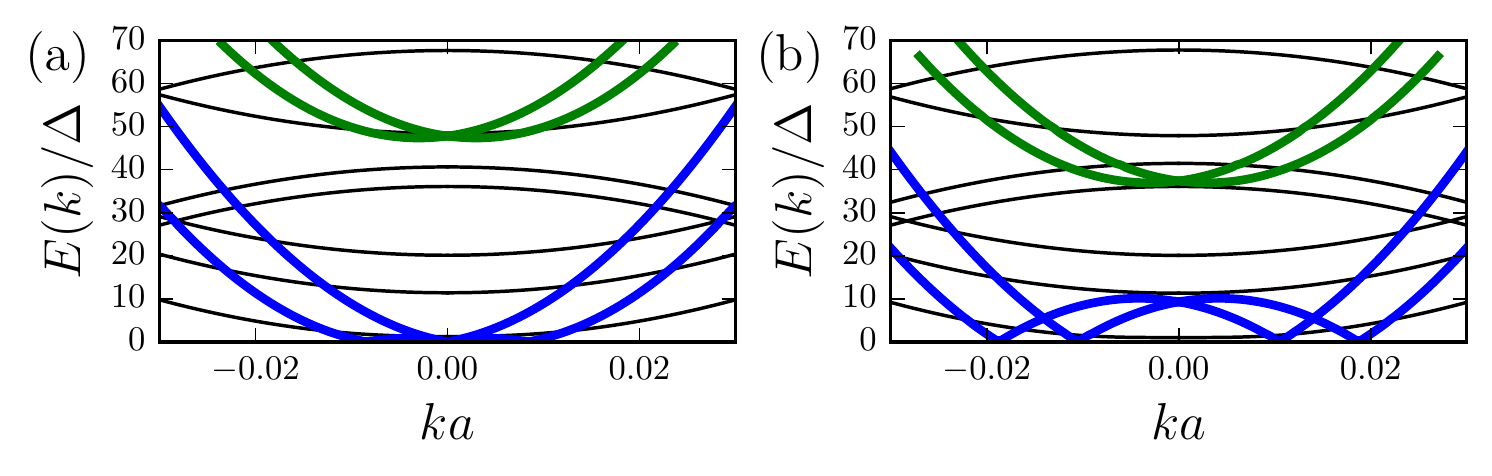}
\caption{\label{specweak} (a) Excitation spectrum in the absence of tunneling ($t=0$) obtained numerically from  Eq.~\eqref{Htot}. The two lowest subbands of the wire are distinguished by color, and black curves correspond to subbands of the superconductor. (b) If a weak tunnel coupling is turned on ($t/t_s=0.035$), the subbands of the wire undergo a substantial shift in energy ($\delta E_n\sim10\Delta$). Tight-binding parameters are fixed to $d/a=42$, $W/a=175$, $\mu_s/t_s=0.1$ \cite{tb}, $\Delta/t_s=10^{-4}$, $t_w/t_s=5$, $\mu_w=0$, $\alpha/(at_s)=0.05$, $\Delta_Z=0$.}
\end{figure}

First, we study the transition from the weak-coupling regime, characterized by a proximity-induced gap $E_g\ll\Delta$, to the strong-coupling regime, characterized by $E_g\sim\Delta$, focusing on the behavior of subbands that originate in the wire (\emph{i.e.}, those subbands that have zero weight in the superconductor in the limit $t=0$) as a function of tunneling strength $t$. We obtain the spectrum $E(k)$ numerically from  Eq.~\eqref{Htot}. As the Fermi wavelength of the superconductor is much smaller than that of the semiconductor, the spectrum consists primarily of subbands originating in the superconductor; wire subbands are distinguished by their appreciable spin-orbit splitting $E_{so}$ and small effective mass $m^*$ (see for example Fig.~\ref{specweak}).

In the absence of tunneling [Fig.~\ref{specweak}(a)], the spectrum of the wire at $k=0$ is given by
\begin{equation} \label{spectrum}
E_n(0)=E_1(0)+\frac{\hbar^2\pi^2}{2m^*W^2}(n^2-1),
\end{equation}
where $n\in\mathbb{Z}^+$ is a subband index and $E_n(0)$ is the energy of the $n$th subband at $k=0$. In the presence of a weak tunnel coupling [Fig.~\ref{specweak}(b)], the superconductor induces a small gap and, even more strikingly, a very substantial energy shift on the subbands of the wire. We define the band shift of the $n$th subband of the wire, which is a function of tunneling $t$, as
\begin{equation}
\delta E_n=|E_n^t (0)-E_n^{t=0}(0)|.
\end{equation}
In the weak-coupling limit of Fig.~\ref{specweak}(b), the band shift $\delta E_n\sim10\Delta$ is more than two orders of magnitude larger than the induced proximity gap ($E_g\sim0.03\Delta$).

\begin{figure}[t!]
\includegraphics[width=\linewidth]{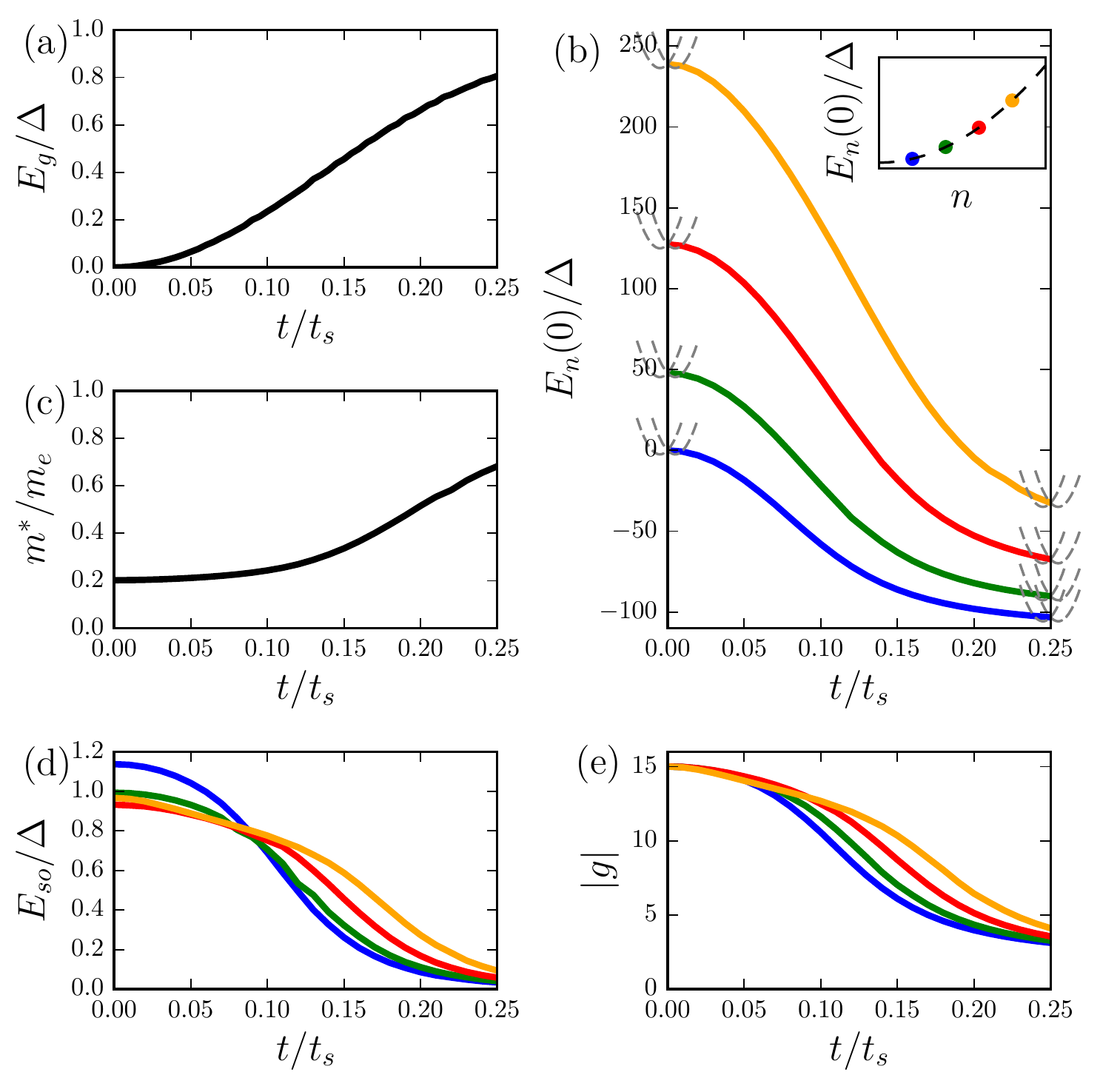}
\caption{\label{specvst} 
(a) Proximity-induced gap $E_g$ increases as a function of tunneling strength $t$, with a large gap being induced only for $t\sim t_s$. (b) The four lowest transverse subbands of the wire (distinguished by color) shift in energy as $t$ is increased. The energy of each subband is measured at $k=0$, as schematically indicated by gray dashed subbands, and the energy of occupied bands is negative. Inset: Subband energy increases quadratically with index $n$ (evaluated at $t=0.25t_s$), in agreement with Eq.~\eqref{spectrum}. (c) The effective mass $m^*$, which is obtained by fitting panel (b) to Eq.~\eqref{spectrum}, increases as a function of $t$. In the limit of large $t$, the mass approaches that of the superconductor ($m^*/m_e\to1$). (d-e) The spin-orbit splitting $E_{so}$ and $g$-factor $|g|$ of each subband are reduced as a function of $t$, also approaching their values in the superconductor ($E_{so}\to0$ and $|g|\to2$) in the limit of large $t$. All parameters are the same as in Fig.~\ref{specweak} [with $|g|=15$ in panel (e)].}
\end{figure}

The evolution of the wire spectrum from the weak- to the strong-coupling limit as a function of $t$ is shown in Fig.~\ref{specvst}. To reach the limit of strong coupling (defined such that $E_g\sim\Delta$), the tunneling strength must be made comparable to $t_s$ [see Fig.~\ref{specvst}(a)]; therefore, a substantial gap $E_g$ can be induced only if there is an extremely high-quality semiconductor/superconductor interface. When such strong tunneling is present, we also observe a very large energy shift in all subbands of the wire [see Fig.~\ref{specvst}(b)], with the bottom of each subband saturating to a different energy at large $t$. This band shift is significantly larger for higher subbands, and, as a result, it requires a larger tunneling strength for higher subbands to reach their saturation positions. Crucially, for all values of $t$, we find that each band is shifted such that Eq.~\eqref{spectrum} remains satisfied [see Fig.~\ref{specvst}(b) inset], provided that we allow the effective mass $m^*$ to acquire a $t$-dependence. The effective mass $m^*$, which can be found by fitting Fig.~\ref{specvst}(b) to Eq.~\eqref{spectrum}, increases as a function of $t$ as the bands originating from the semiconductor acquire a larger weight inside the superconductor [Fig.~\ref{specvst}(c)]. Additionally, the spin-orbit splitting $E_{so}$ [Fig.~\ref{specvst}(d)] and $g$-factor [Fig.~\ref{specvst}(e)] (extracted from the Zeeman splitting at $k=0$ \cite{gfactor}) of each subband are reduced as a function of $t$. All parameters of the semiconducting wire saturate to their corresponding values within the superconductor in the limit of large $t$ ($m^*/m_e\to1$, $E_{so}\to0$, and $|g|\to2$). However, similarly to the band shifts, this parameter renormalization is not the same for all subbands, as higher subbands require a larger tunneling strength to become fully renormalized.

\subsection{Role of finite width $W$} \label{secFiniteW}
\begin{figure}[t!]
\centering
\includegraphics[width=\linewidth]{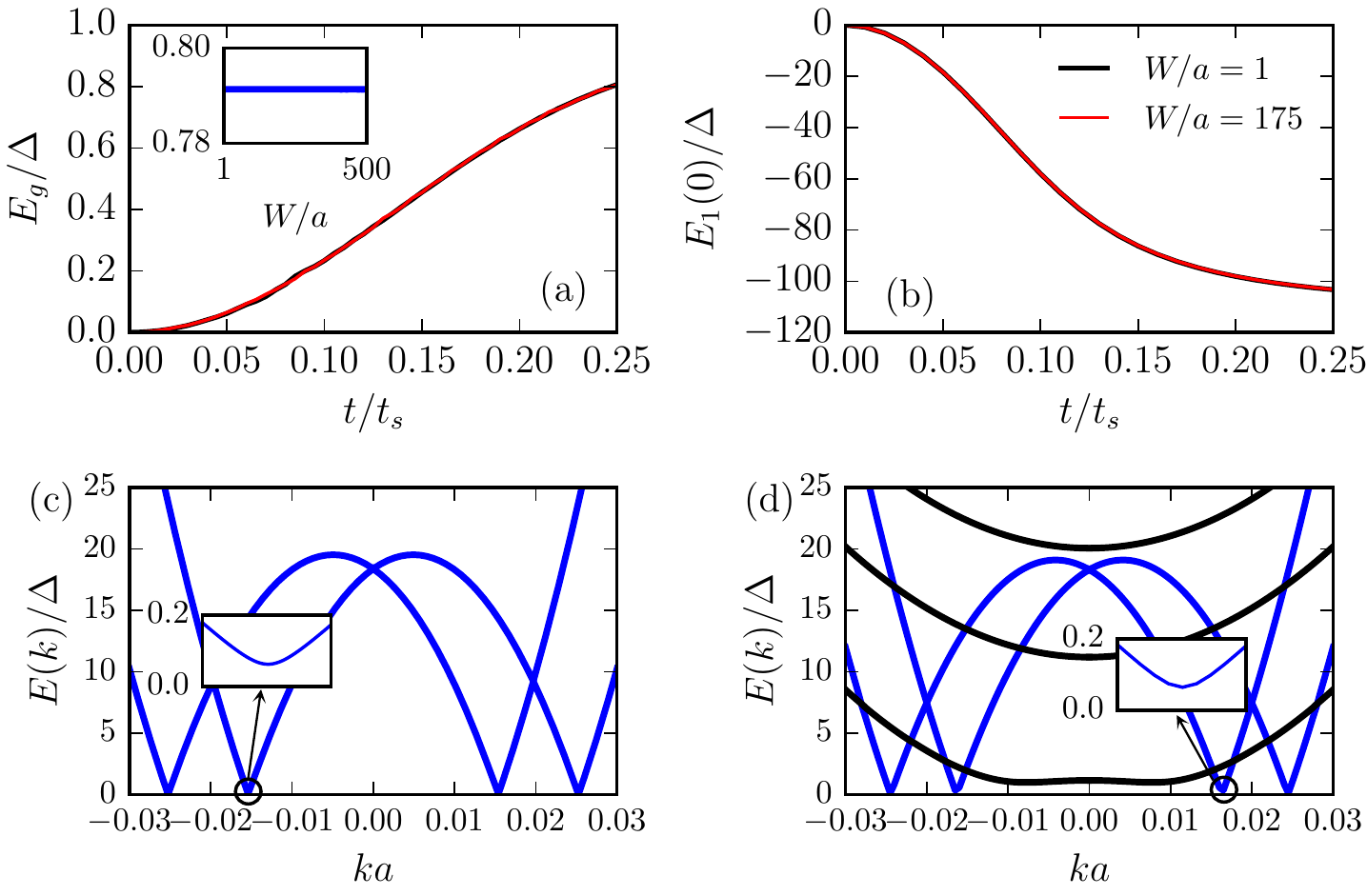}
\caption{\label{finiteW} (a) The induced gap $E_g$ is the same for both widths $W/a=1$ (black) and $W/a=175$ (red) at all values of $t$. Inset: $E_g$ (blue) is independent of width $W$ over several orders of magnitude. (b) Energy of lowest subband at $k=0$, $E_1(0)$, is also the same for $W/a=1$ and $W/a=175$ at all values of $t$. (c-d) Excitation spectrum for $W/a=1$ and $W/a=175$, respectively. The spectrum of the lowest wire subband $E_1(k)$ (blue) is virtually unchanged as the width $W$ is increased and does not couple to low-energy superconducting subbands (black) appearing for $W/a=175$. All parameters are the same as in Fig.~\ref{specweak}.}
\end{figure}

To elucidate the dependence of the spectrum on the finite width $W$, we present a comparison of the cases $W/a=1$ (henceforth referred to as a 2D geometry) and $W/a\gg1$ (henceforth referred to as a 3D geometry) in Fig.~\ref{finiteW}. We find that both the induced gap $E_g$ [Fig.~\ref{finiteW}(a)] and the energy of the lowest transverse subband at $k=0$, $E_1(0)$ [Fig.~\ref{finiteW}(b)], are identical in the 2D and 3D cases. In fact, we find that $E_g$ is completely independent of $W$ over several orders of magnitude [Fig.~\ref{finiteW}(a) inset], suggesting that the width $W$ plays a rather trivial role in the proximity effect. To better understand these results, we plot the spectrum explicitly in Fig.~\ref{finiteW}(c-d). In the 3D limit, despite there being several transverse superconducting subbands at low energies (and thus a significantly reduced level spacing in the superconductor), these subbands do not couple to the lowest subband of the wire, as evidenced by the absence of anticrossings in the spectrum. As a result, the spectrum of the lowest wire subband is virtually unchanged as the width $W$ is increased. We provide a detailed analytical justification of this numerical result in Appendix~\ref{AppendixA}.

The fact that the spectrum of the lowest wire subband is independent of $W$ can be understood in a rather simple way. In the limit $W\to\infty$, it is possible to define an additional conserved momentum $k_x$. Therefore, the spectrum in the limit $W\to\infty$ is the same as that for $W\to0$, with the simple replacement $k\to\sqrt{k^2+k_x^2}$. As the spectrum is identical for $W\to\infty$ and $W\to0$, it is unsurprising that it is independent of $W$ for intermediate widths as well. Note that such an argument cannot be made if we take $d\to\infty$, as the semiconductor/superconductor interface always breaks translational invariance in the $z$-direction.

Based on Fig.~\ref{finiteW}, we conclude that the finite width $W$ of the system only introduces a finite level spacing between transverse subbands but otherwise has no effect on the induced gap or band shift. It is very computationally expensive to treat the finite width $W$ explicitly as we have done to this point, thus, we forego doing so in the calculations that follow. A 2D calculation can be performed to reliably reproduce the behavior of the lowest subband, and by calculating the effective mass in this subband (for a given $t$), we can deduce the transverse level spacing in the 3D limit. The only drawback to using such an approach is that it does not fully capture the weaker parameter renormalization in higher transverse subbands, as shown in Fig.~\ref{specvst} and as discussed in Sec.~\ref{secStrongCoupling}. However, this is a relatively minor omission and should not affect any of our results qualitatively.

\subsection{Effect of Disorder} \label{secDisorder}
So far we have considered only clean translationally invariant systems; however, the superconductor that is utilized in any realistic experimental setup will inevitably be disordered. As discussed in {Sec.~\ref{secIntro}}, the breaking of translational symmetry by disorder could allow a stronger proximity coupling between the wire and superconductor at lower energies (due to the fact that the superconductor has several occupied subbands at the Fermi level). In this section, we study the influence of various types of disorder on the induced gap in the wire, thereby relaxing the requirement of momentum conservation imposed in Eq.~\eqref{Htot}. While moderate chemical potential disorder within the superconductor has very little effect on the proximity gap, we find that both interface inhomogeneity as well as strong surface disorder lead to a small enhancement of the proximity gap. For computational reasons, all disorder calculations are performed in a 2D geometry ($W/a=1$) and without SOI ($\alpha=0$).

\subsubsection{Disorder in superconductor} \label{secBulkDisorder}
\begin{figure}[t!]
\centering
\includegraphics[width=\linewidth]{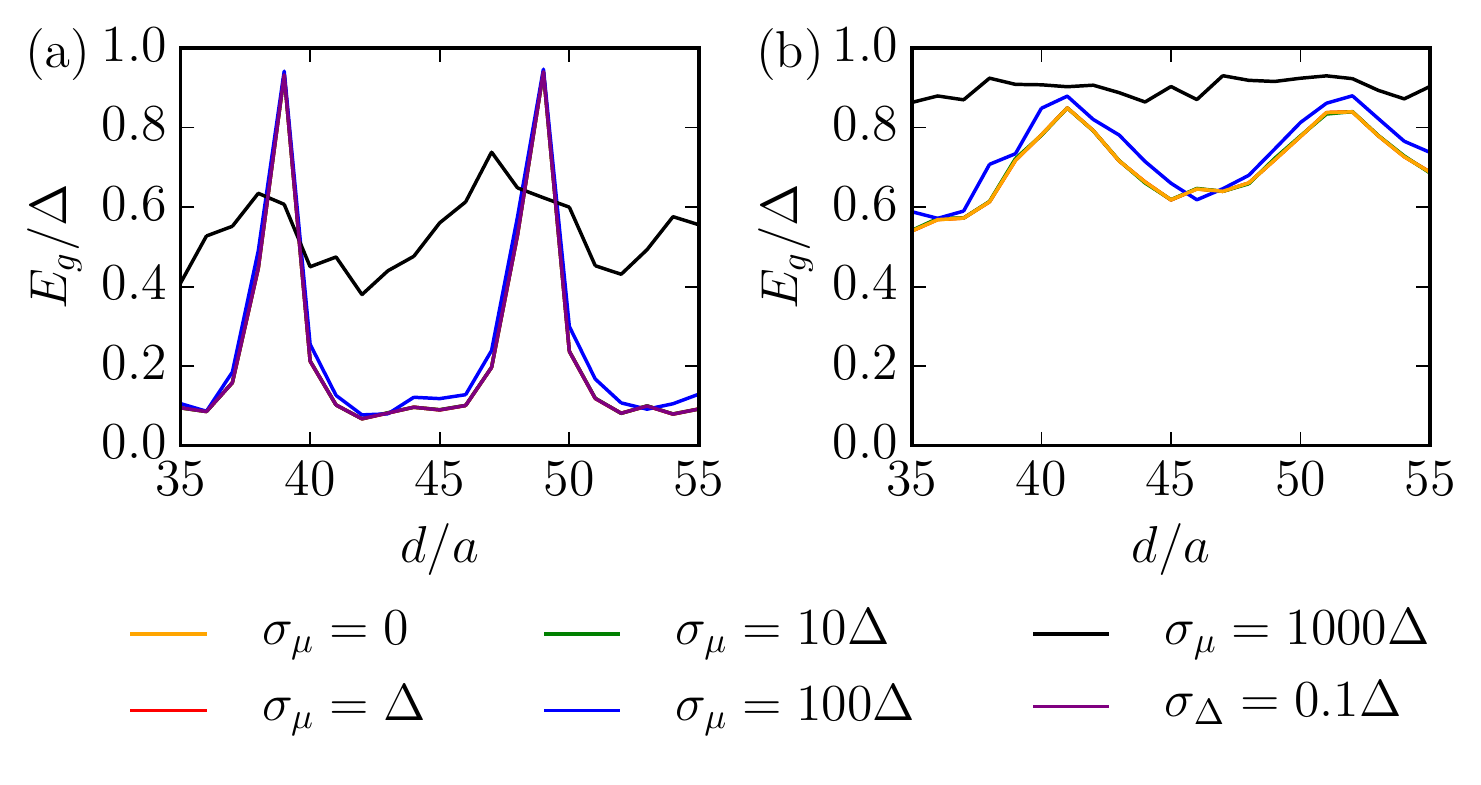}
\caption{\label{disorder} Induced gap $E_g$ as a function of superconductor thickness $d$ in a disordered 2D tight-binding model (with length $L/a=3\times10^4$) with various disorder strengths $\sigma_{\mu(\Delta)}$, plotted for (a) $t=0.05t_s$ and (b) $t=0.25t_s$. In both the weak- and strong-coupling limits, the gap is largely unaffected by disorder unless chemical potential fluctuations in the superconductor are extreme ($\sigma_\mu=1000\Delta=\mu_s$). Tight-binding parameters are fixed to $d/a=42$, $W/a=1$, $\mu_s/t_s=0.1$, $\Delta/t_s=10^{-4}$, $t_w/t_s=5$, $\mu_w/\Delta=1$, $\alpha=0$, $\Delta_Z=0$.}
\end{figure}

First, we incorporate disorder within the superconductor as random on-site Gaussian-distributed fluctuations in the chemical potential $\mu_s$ and pairing potential $\Delta$, $\mu_s\to\mu_s+\delta\mu_s$ and $\Delta\to\Delta+\delta\Delta$. Fluctuations are taken to have standard deviation $\sigma_{\mu(\Delta)}$ and zero mean, $\langle\delta\mu_s\rangle=\langle\delta\Delta\rangle=0$. The wire is taken to be clean. Furthermore, we consider a finite length $L$ of our system chosen such that, in the absence of disorder, we reproduce the proximity gap previously obtained in the $L\to\infty$ limit in which the momentum $k$ is conserved. We find that the induced gap is largely unaffected by moderate disorder in both the weak-coupling [Fig.~\ref{disorder}(a)] and strong-coupling [Fig.~\ref{disorder}(b)] limits. The gap is enhanced and the sharp interference peaks, which arise due to the finite thickness of the superconducting layer \cite{Reeg:2017_3}, are smeared only when fluctuations in the chemical potential become comparable to the chemical potential itself, $\sigma_\mu\sim\mu_s$.

The fact that the well-pronounced interference peaks [see Fig.~\ref{disorder}] survive in the dirty limit can be understood straightforwardly on physical grounds. Due to the large mismatch in effective mass and Fermi momentum between the wire ($k_{Fw}$) and superconductor ($k_{Fs}$), the relevant superconducting trajectories that are responsible for inducing a gap have momentum along the wire axis $k\lesssim k_{Fw}\ll k_{Fs}$. However, these trajectories are nearly perpendicular to the semiconductor/superconductor interface and are ballistic, since typical values for the mean free path $\ell$ of (bulk) Al are larger than the thickness of the superconducting film ($\ell\gtrsim d$).
More quantitatively, it was shown in Ref.~\cite{Reeg:2017_3} that the relevant energy scale determining the tunneling strength needed to induce a sizable gap in the wire in the clean limit is the level spacing of the superconducting layer, $\pi\hbar v_{Fs}/d$. On the other hand, in the dirty limit the relevant scale is given by the Thouless energy $\hbar D/d^2$ \cite{Usadel:1970,Belzig:1996,Belzig:1999,Reeg:2014}, where $D\sim v_{Fs}\ell$ is the diffusion coefficient. However, as $\ell\sim d$, the two energy scales are comparable ($\hbar v_{Fs}/d\sim\hbar D/d^2$). Disorder therefore does not qualitatively change the behavior of the induced gap by introducing a low-energy scale unless $d\gg\ell$. The bulk limit of the superconductor, where the induced gap $E_g$ no longer depends on $d$, is reached only for $d\gg\xi_\text{dirty}$ (or, equivalently, $\hbar D/d^2\ll\Delta$), where $\xi_\text{dirty}=\sqrt{\ell\xi_\text{clean}}$ is the effective coherence length of the superconductor in the dirty limit \cite{bulkdisorder}. We note that these physical arguments do not rely on the width $W$ of the system being negligibly small, and hence we expect them to hold also in the 3D limit.

\subsubsection{Disorder in tunneling}
Next, we incorporate possible interface inhomogeneity through fluctuations in the tunneling strength $t\to t+\delta t$ (which again are Gaussian distributed with standard deviation $\sigma_t$ and zero mean, $\langle\delta t\rangle=0$). As shown in Fig.~\ref{disorder2}(a), fluctuations in the tunneling amplitude lead to an increase in the induced gap. This is a reflection of the finite level spacing within the superconductor. When the length $L$ of the system is finite, the momentum along this direction becomes quantized. If the tunnel barrier is uniform along the interface between the two materials, then only subbands in the wire and superconductor with the same quantum number can couple (see also discussion in Appendix~\ref{AppendixA}), but inhomogeneity can lead to nonzero matrix elements between states with different quantum numbers and, hence, an increase in the gap of the wire. However, as we observe, interface fluctuations must be comparable to the tunneling strength in order to induce a qualitative change to the behavior of the gap in the clean case; in the epitaxial Al devices, which were developed specifically to have a very homogeneous interface, this seems an unlikely scenario.

\subsubsection{Strong surface disorder}

\begin{figure}[t!]
\centering
\includegraphics[width=\linewidth]{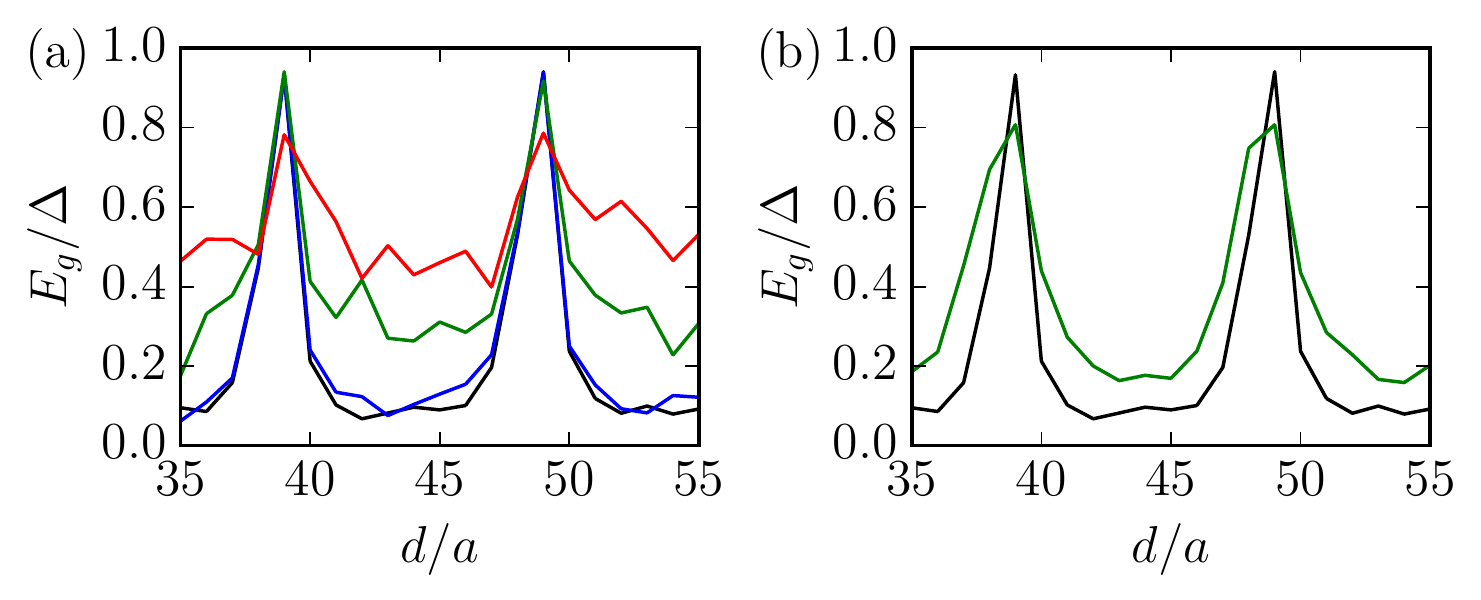}
\caption{\label{disorder2} (a) Induced gap $E_g$ for various strengths of tunneling fluctuations: $\sigma_t=0$ (black), $\sigma_t=0.2t$ (blue), $\sigma_t=t$ (green), and $\sigma_t=2t$ (red). The gap is largely unaffected unless fluctuations become comparable to tunneling strength $t$. (b) Strong surface disorder (green) slightly enhances the induced gap compared with the clean limit (black). Surface disorder is incorporated through chemical potential fluctuations with $\sigma_\mu=1000\Delta$ on the five sites furthest from the interface and with $\sigma_\mu=10\Delta$ on the remaining sites within the superconductor. Tight-binding parameters are the same as in Fig.~\ref{disorder}.}
\end{figure} 

Finally, we investigate the effects of strong surface disorder of the superconducting layer, which could be present due to an oxidized surface layer. We model this scenario by taking very large chemical potential fluctuations on the five sites furthest from the interface and moderate fluctuations on the remaining sites. We find that surface disorder can modestly enhance the size of the induced gap away from resonance and broaden the resonance peak [see Fig.~\ref{disorder2}(b)]. This behavior can be understood in the following way. As explained previously, in the clean limit only trajectories within the superconductor with $k\ll k_{Fs}$ can open a gap in the wire. In the presence of strong surface scattering, trajectories that begin with momentum $k\sim k_{Fs}$ can scatter at the surface into trajectories with $k\ll k_{Fs}$. Therefore, strong surface scattering allows for more trajectories to open a gap and the magnitude of the gap is increased. However, this rough surface essentially sets an upper bound on the mean free path, such that ${\ell}\sim d$. As explained in {Sec.~\ref{secNumerics}\ref{secBulkDisorder}}, these values of $\ell$ are not expected to have a substantial effect on the induced gap.

\section{Experimental Consequences} \label{secEstimates}
In Sec.~\ref{secNumerics}, we argued that a 3D geometry with various types of disorder present can actually be very well described by a clean 2D model. The elimination of the finite width $W$ allows us to explore a much larger region of parameter space in a tight-binding calculation. In this section, we explicitly model the experimental setup of an InAs 2DEG coupled to an epitaxial Al layer of thickness $d\sim10$ nm \cite{Nichele:2017,Suominen:2017}. We also provide a discussion of the feasibility of utilizing such a setup to realize a topological phase. In addition, we propose an experimental test of our theory.

\subsection{Proximity-induced gap and band shift} \label{sec2D}

As a starting point for our calculation, we note that all proximity-induced gaps that have been observed in epitaxial systems are a sizable fraction of the Al gap, $E_g\sim\Delta$ \cite{Gazibegovic:2017,Zhang:2017_2,Kjaergaard:2016,Shabani:2016,Kjaergaard:2017,Suominen:2017,Nichele:2017,Vaitiekenas:2017,Deng:2017}. We thus assume that the system is in the strong-coupling limit; \emph{i.e.}, that the tunneling strength $t$ is large enough such that a sizable gap is induced for all $d\sim10$ nm.


\begin{figure}[t!]
\centering
\includegraphics[width=\linewidth]{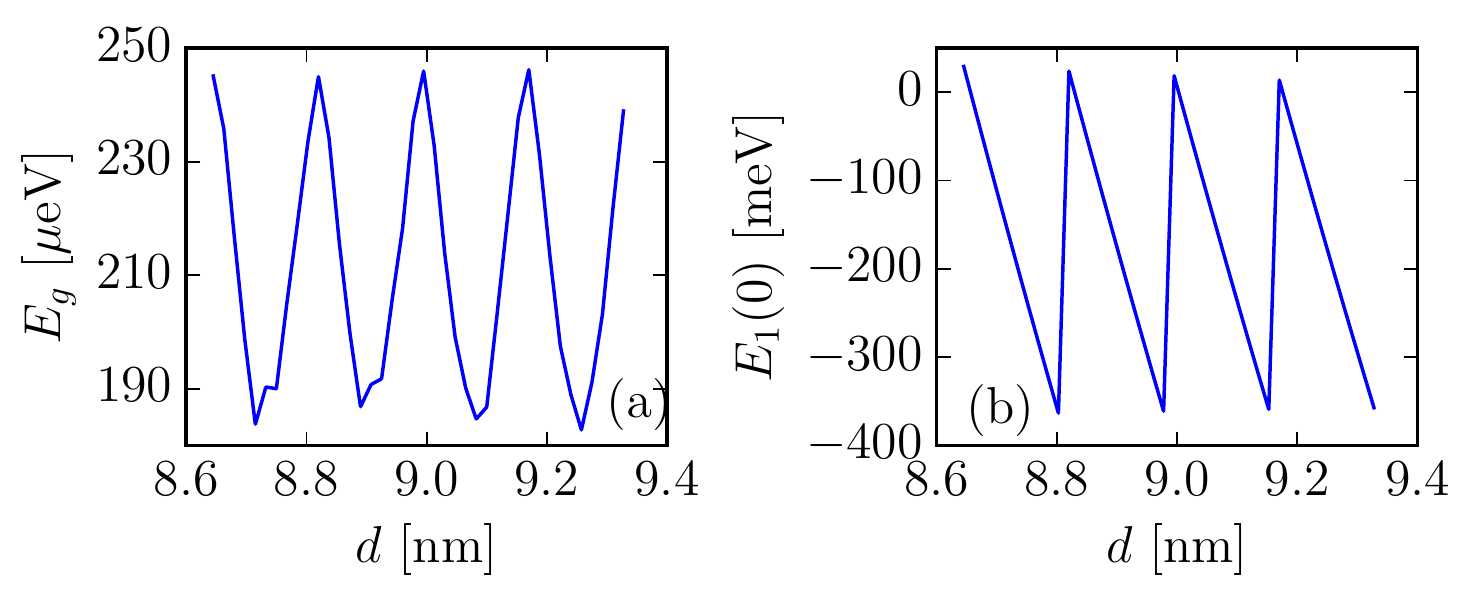}
\caption{\label{deltamuvsd} (a) Proximity-induced gap $E_g$ and (b) energy of lowest subband at $k=0$, $E_1(0)$, for parameters corresponding to epitaxial Al/InAs. The tunneling strength $t=0.1t_s$ is chosen large enough such that a sizable gap $E_g$ is induced for all values of $d$. In the strong-coupling limit, the wire subband undergoes a huge band shift $\delta E_1=|E_1(0)|\sim200$ meV. The tight-binding parameters are set to $t_s=117$ eV (corresponding to lattice spacing $a=0.175$ \AA), $\mu_s=11.7$ eV, $\Delta=250$ $\mu$eV, $t_w=50t_s$ (corresponding to $m^*=0.02m_e$), $\mu_w=0$, $\alpha=0.42$ eV \AA (corresponding to $E_{so}=250$ $\mu$eV), and $\Delta_Z=0$.}
\end{figure}

When $t$ is made large enough to satisfy the strong-coupling condition [see Fig.~\ref{deltamuvsd}(a)], we find that the wire subband undergoes a huge energy shift. Consistent with the analytical results of Ref.~\cite{Reeg:2017_3}, the magnitude of this band shift is comparable to the level spacing in the superconducting layer, $\pi\hbar v_F^\text{Al}/d\sim500$ meV (with $v_F^\text{Al}=2\times10^6$ m/s), and is very sensitive to the thickness $d$, varying between $E_1(0)\in(-400\text{ meV}, 50\text{ meV})$ with a period that is half of the Fermi wavelength of Al ($\approx2$ \AA) [see Fig.~\ref{deltamuvsd}(b)]. Additional results of this calculation are provided in Appendix~\ref{AppendixB}. Unsurprisingly, the large band shift in the strong-coupling limit is accompanied by significant renormalizations of the effective mass ($m^*\sim0.3m_e$), spin-orbit splitting ($E_{so}\sim10$ $\mu$eV), and $g$-factor ($|g|\sim2$).

\subsection{Metallization and impact on topological superconductivity} \label{secTopological}

\begin{figure}[t!]
\centering
\includegraphics[width=\linewidth]{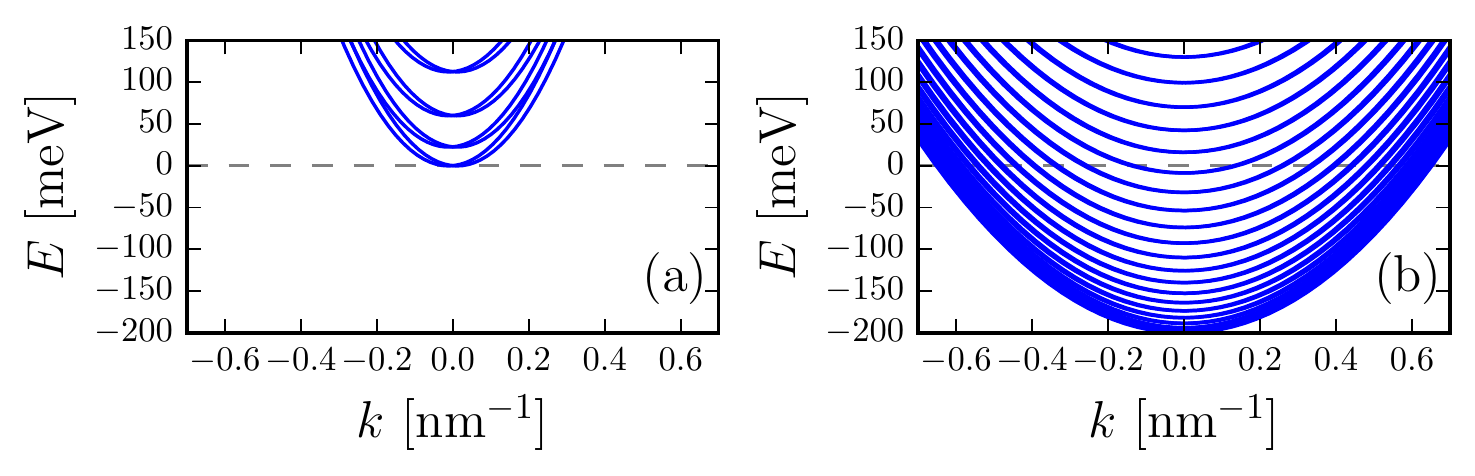}
\caption{\label{renormalization} Schematic illustration of the metallization of the Rashba wire. (a) Spectrum of the wire in the absence of tunneling. We assume the chemical potential to be tuned to the crossing point of the lowest subband and take the wire to have a small effective mass $m^*\sim0.02m_e$ and large spin-orbit splitting $E_{so}\sim250$ $\mu$eV. (b) Renormalization of the spectrum in the strong-coupling limit. The lowest wire subband experiences a large band shift [taken from Fig.~\ref{deltamuvsd} to be $\delta E_1\sim-200$ meV], and renormalization of the effective mass ($m^*\sim0.2m_e$) leads to the occupation of many wire subbands. The subbands of the wire also have a significantly reduced spin-orbit splitting $E_{so}\sim25$~$\mu$eV.}
\end{figure}

Using the results of the 2D calculation of Sec.~\ref{sec2D}, we present a schematic illustration of the 3D spectrum renormalization in Fig.~\ref{renormalization}. Due to the extreme sensitivity of the induced gap and band shift on the thickness of the superconducting layer $d$, we are only able to provide a qualitative picture of this spectrum renormalization in a typical device.

We assume that the quantum wire has width $W=50$ nm \cite{Nichele:2017} and that the chemical potential of the bare wire (without the superconducting layer) is tuned to the Rashba crossing point at $k=0$ of the lowest subband (though, as shown in Appendix~\ref{AppendixB}, our final results are independent of this assumption). Taking $m^*=0.02m_e$ for the mass of (bare) InAs, the spectrum at $k=0$ in the absence of tunneling is given by [Fig.~\ref{renormalization}(a)] 
\begin{equation}
E_n(0)=(7.5\text{ meV})(n^2-1).
\end{equation}
In the strong-coupling limit, based on Fig.~\ref{deltamuvsd}, we take an intermediate value for the band shift of $ \delta E_1\sim200$ meV; in this case, the mass is renormalized to $m^*\sim0.2m_e$ (see Appendix~\ref{AppendixB}) and the spectrum at $k=0$ is given by [Fig.~\ref{renormalization}(b)]
\begin{equation} \label{spectrumlarget}
E_n(0)=(-200\text{ meV})+(0.75\text{ meV})(n^2-1).
\end{equation}
Figure \ref{renormalization}(b) illustrates the central finding of our work. Due to the large band shift, which acts as an effective enhancement of the chemical potential of the wire, and increase in effective mass, many additional transverse wire subbands become occupied and the semiconductor is essentially metallized by the superconductor. While there are certainly more electrons in the wire in the metallized limit (compared to the bare wire), as noted previously, these electrons can be supplied to the system by external leads, and the number of electrons added to the wire is negligible compared to the total number of electrons in the system. We also note that states belonging to the spectrum of Fig.~\ref{renormalization}(b) are delocalized throughout the superconductor, thus helping to reduce the redistribution of charge into the wire.

The Zeeman splitting required to drive the system through a topological phase transition is given by
\begin{equation}
\Delta_Z=\sqrt{\mu_\text{min}^2+E_g^2},
\end{equation}
where $\Delta_Z$ includes the renormalized $g$-factor of the wire and $|\mu_\text{min}|=\min[|E_n(0)|]$ is the effective chemical potential of the transverse subband that lies closest to the Fermi level. Therefore, to reach a topological phase in the system, the chemical potential must ideally be tuned to the Rashba crossing point of one of the transverse wire subbands, such that $\mu_\text{min}=0$. In this special case, it is possible to reach a topological phase before destroying superconductivity in the Al shell; taking for example a renormalized $g$-factor of $|g|\sim5$ (see Appendix~\ref{AppendixB}) and an induced gap $E_g\sim200$ $\mu$eV, the topological phase can be reached with a field strength $B\sim1.5$ T, which is smaller than the critical field of a thin Al layer (which is Clogston-limited \cite{Clogston:1962}, $B_c=\Delta/\sqrt{2}\mu_B\sim3$ T). Therefore, the renormalization of the semiconductor alone does not make the topological phase inaccessible \emph{a priori}. However, it is very unlikely that the limit $\mu_\text{min}=0$ will be satisfied in practice; as we have seen, the position of the Fermi level is entirely determined by the large band shift, which is extremely sensitive to $d$ and is therefore highly device dependent. The worst-case scenario corresponds to a maximal band shift $|E_1(0)|\sim400$ meV that places the Fermi level at the midpoint between the two closest transverse subbands. For a maximal band shift, the transverse subband spacing in the vicinity of the Fermi level can be estimated as $(\pi\hbar/W)\sqrt{2|E_1(0)|/m^*}\sim35$ meV, thus placing an upper bound of $|\mu_\text{min}|\sim17.5$ meV. In this case, the field strength that would be required to reach a topological phase is $B=2|\mu_\text{min}|/|g|\mu_B\sim120$ T. Of course, such an unrealistically large value for the required magnetic field simply means that superconductivity in Al will be destroyed before reaching the topological phase \cite{fielddependence}. Due to the relative lack of control over the band shift in the limit of a thin superconducting layer, the field strength required to reach a topological phase can lie anywhere in the range between 1.5 T and 120 T with roughly equal probability. Given that superconductivity in the Al layer is destroyed at $B_c\sim3$ T, it is thus very challenging to reach the topological phase in such a device. In order to reliably produce a topological phase, one needs to be able to control the chemical potential over a range of $\sim10-20$ meV in order to offset large $|\mu_\text{min}|$ \cite{orbital}. While current experiments on 2DEGs \cite{Suominen:2017,Nichele:2017} do not have gates available to tune the chemical potential, even if such gates were implemented we expect that screening effects arising from the strong coupling and close proximity to a metal will not allow for such a large range of tunability.

Nevertheless, even if a subband is shifted to the ideal position $\mu_\text{min}=0$, we must stress that this is no guarantee that one will observe well-separated Majorana fermions in the system. First, the spin-orbit splitting is renormalized to a prohibitively small value $E_{so}\sim25$ $\mu$eV. As it is the SOI that is responsible for inducing $p$-wave pairing in the wire, the localization of the Majorana wave function could (possibly greatly) exceed the length of the wire. Second, it is unclear whether 1D topological physics would be observed in the metallized case due to the presence of many occupied transverse levels.

We note that our main result, namely the metallization of the wire depicted in Fig.~{\ref{renormalization}}, is independent of our choice $\mu_w=0$. We show in Appendix~{\ref{AppendixC}} that the spectrum in the strong-coupling limit is independent of $\mu_w$. Hence, our result holds even if there are several occupied transverse subbands at $t=0$. Additionally, we checked that our results also hold in the presence of Fermi surface anisotropy within the superconductor, which we implement by allowing for anisotropic hopping $t_{s,z}=2t_{s,y}$.

\subsection{Controlling the band shift}

\begin{figure}[t!]
\centering
\includegraphics[width=\linewidth]{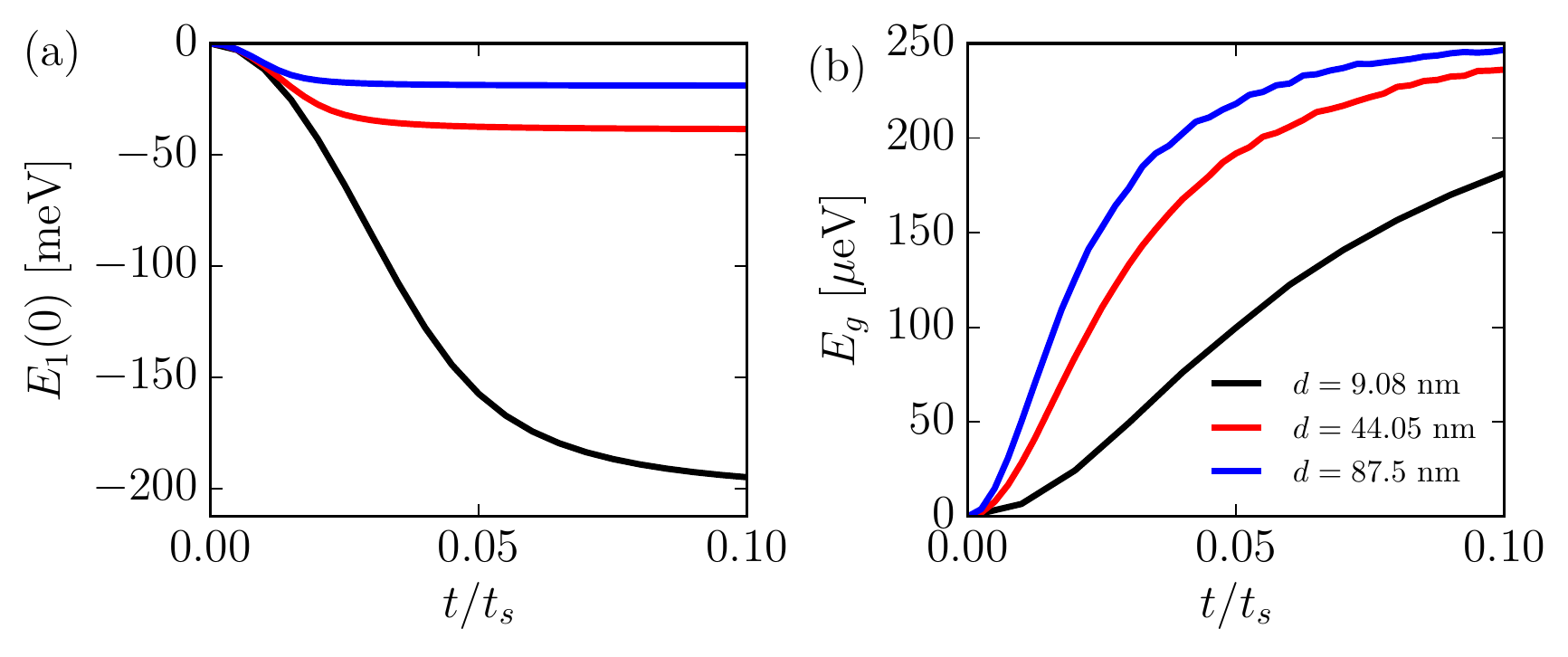}
\caption{\label{shiftlarged} (a) The band shift is significantly reduced by increasing the thickness $d$. (b) When $d$ is increased, the crossover from weak-coupling ($E_g\ll\Delta$) to strong-coupling ($E_g\sim\Delta$) occurs at much smaller $t$. Tight-binding parameters are the same as in Fig.~\ref{deltamuvsd}.}
\end{figure}

As suggested in Ref.~\cite{Reeg:2017_3}, the detrimental band shift can be reduced by increasing the thickness of the superconducting layer. More specifically, while the band shift still exhibits oscillations on the scale of the Fermi wavelength as in Fig.~\ref{deltamuvsd}(b), the oscillation amplitude and typical magnitude (i.e., the average over a single oscillation) are reduced with increasing $d$ \cite{Reeg:2017,Schrade:2017}.  We confirm this numerically, as shown in Fig.~\ref{shiftlarged}(a). Therefore, the size of the band shift can be tuned experimentally by varying the thickness $d$ of the superconducting layer, and measuring a sharp decrease in the band shift (for example, by angle-resolved photoemission spectroscopy) in systems with larger thickness would constitute a clear experimental verification of our theory. The band shift could also be controlled through the addition of a tunnel barrier between the superconductor and 2DEG, with the band shift decreasing in magnitude as the thickness of the barrier layer is increased.

Even though increasing the thickness $d$ of the superconducting layer can reduce the band shift, this is not necessarily beneficial for inducing a topological phase. As shown in Fig.~\ref{shiftlarged}(b), increasing $d$ also shifts the crossover from weak-coupling ($E_g\ll\Delta$) to strong-coupling ($E_g\sim\Delta$) to significantly smaller $t$. The tunneling strength is a property of the interface, so it should not be affected by the thickness of the superconducting layer. Therefore, if tunneling is strong enough to induce a sizable gap for $d\sim10$ nm as it is for the epitaxial interface, the system will be deep within the strong-coupling regime if the thickness $d$ is increased. In this limit, all semiconducting properties are completely eliminated by the strong coupling to the superconductor and it is challenging to realize a topological phase \cite{Potter:2011}.

\section{Conclusion} \label{secConclusion}
We have studied the proximity effect in a quasi-1D quantum wire (defined within a 2DEG) strongly coupled to a thin disordered superconducting layer. We showed that, even in the strong-coupling limit, the behavior of the lowest transverse subband in such a system can be very well described by a single 1D channel coupled to a clean 2D superconductor, as studied analytically in Ref.~\cite{Reeg:2017_3}. Utilizing this result, we found that if the proximity-induced gap in an epitaxial Al/InAs heterostructure is comparable to the gap of Al (as observed experimentally), the semiconductor is metallized by the superconductor. Not only do the subbands of the wire undergo a huge band shift $\sim200$ meV, which leads to the occupation of many transverse levels and effectively places the wire far from the 1D limit, but the semiconducting properties that are attractive for realizing a topological phase (large $g$-factor, large spin-orbit splitting, small effective mass) are also significantly renormalized toward their metallic values. We argued that this metallization effect makes it challenging to realize a topological phase in an epitaxial Al/InAs setup, with the ability to do so being largely device dependent. We also proposed that our theory can be verified experimentally by observing a decrease in the magnitude of the band shift when the thickness $d$ of the superconducting layer is increased.

Despite the recent emphasis on electron-electron interaction effects inside hexagonal nanowires \cite{Antipov:2018,Woods:2018,Mikkelsen:2018}, we do not consider such effects in our model. Most importantly, interaction effects give rise to a nontrivial spatial profile of the electrostatic potential across the diameter of such nanowires, which spans roughly $50-100$ nm ({\it i.e.}, interactions give rise to a band bending effect within the semiconductor). In the setup that we consider, where a quantum wire is defined within a 2DEG, there is no spatial extent over which such a profile can develop (the thickness of the 2DEG is only $\sim5$ nm \cite{Shabani:2016}). Additionally, as the states in the wire are in such close proximity to a metal, one expects that interactions in the wire are heavily screened \cite{Woods:2018}. We also neglect potential Luttinger liquid effects that can suppress the induced superconducting gap in both clean and disordered nanowires \cite{Gangadharaiah:2011,Stoudenmire:2011,Disorder_RG}. It is worth noting that Ref.~\cite{Antipov:2018} suggests that in the strong-coupling limit states in the wire are highly localized near the interface; thus, our model may be applicable to the hexagonal nanowire case as well. However, this result was obtained by treating the superconductor simply as a boundary condition in the Poisson equation~\cite{Antipov:2018}; as pointed out in Ref.~\cite{Woods:2018}, such a treatment does not adequately describe the strong-coupling limit where the states in the nanowire are strongly affected by the presence of the superconductor (for example, the significant reduction in the transverse level spacing of the wire by the proximity effect, one of the key results of our work, is not captured).

We find experimental support for our theory in Refs.~\cite{Deng:2016,Gazibegovic:2017,Zhang:2017_2}. Possible evidence of Majorana fermions in the form of zero-bias conductance peaks have been experimentally observed in both the nanowire \cite{Deng:2016,Zhang:2017_2,Vaitiekenas:2017,Deng:2017} and 2DEG \cite{Suominen:2017,Nichele:2017} epitaxial geometries. In both cases, these zero-bias peaks emerge at finite magnetic field strength from coalescing Andreev bound states that originate from quantum dot-like normal sections at the system ends. For this reason, there has been a significant debate over how to differentiate between zero-bias peaks arising from such Andreev bound states and from Majorana fermions \cite{Deng:2017,Zhang:2017_2,Nichele:2017,Setiawan:2017,Liu:2017,Hell:2017,Rosdahl:2018}. However, interestingly, it has been observed that in the absence of any quantum dot physics, where there are no Andreev bound states at low field strengths, there is no topological gap-closing transition or zero-bias peak emergence with increasing field strength before superconductivity in Al is destroyed \cite{note}. Although this is not definitive confirmation of our model description, it is consistent with the chemical potential being tuned between transverse subbands of the wire (and thus outside of the regime for which topological superconductivity can be induced), which we expect to be the case in a majority of devices.


The metallization of the semiconductor discussed in the present work is a direct consequence of the extremely high-quality interface  provided by the epitaxial growth of Al on InAs. In order to more reliably induce 1D topological superconducting phases, a weaker proximity effect should be sought to a superconductor with a larger gap such as NbTi, which has a gap $\Delta\sim2$ meV that is an order of magnitude larger than that of Al. 



\acknowledgments
We thank M. Leuenberger, C. Marcus, D. Maslov, F. Nichele, J. Nyg\aa rd, and Y. Volpez for helpful discussions. This work was supported by the Swiss National Science Foundation and the NCCR QSIT.

\appendix
\section{Induced gap independent of width} \label{AppendixA}
In Sec.~\ref{secFiniteW}, we showed numerically that the proximity-induced gap is independent of the width $W$ of the quasi-1D quantum wire. In this section, we support our numerical calculations analytically by determining the induced gap within second-order perturbation theory in the weak-coupling limit. In the tunneling-Hamiltonian approach, the tunneling amplitude between a given subband of the wire and a given subband of the superconductor is
\begin{equation}
t=\int_{-d_w}^ddx\int_0^Wdz\,\psi^*_w(x,z)V(z)\psi_s(x,z),
\end{equation}
\begin{figure}[t!]
\centering
\includegraphics[width=\linewidth]{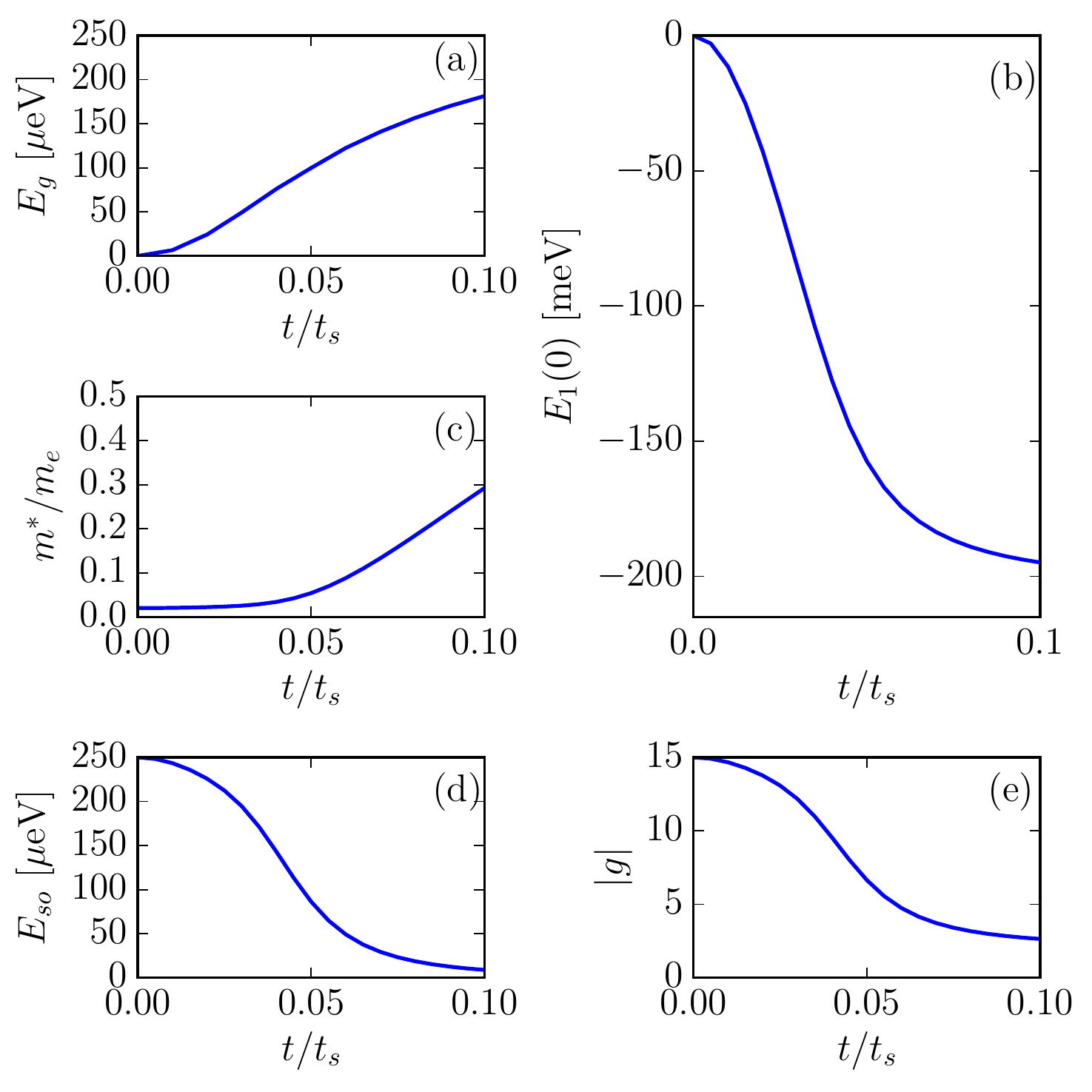}
\caption{\label{real2D} (a) Proximity-induced gap $E_g$, (b) energy of lowest subband at $k=0$, $E_1(0)$, (c) effective mass $m^*$, (d) spin-orbit splitting $E_{so}$, and (e) $g$-factor plotted as a function of tunneling strength $t$. Tight-binding parameters are the same as those in Fig.~\ref{deltamuvsd}, with $d=9.08$ nm (corresponding to $d/a=519$).}
\end{figure}
where $d_w$ is the finite thickness of the wire, $\psi_{w(s)}(x,z)$ is the wave function of a given subband in the wire (superconductor), and $V(z)$ is a barrier potential that we assume is uniform along the interface. [Note that this is not the same $t$ that was introduced in Eq.~\eqref{Ht}.] Given that the wave functions are separable in the coordinates $(x,z)$, the integral over $z$ simply yields the tunneling amplitude $t_0$ in the limit $W=0$,
\begin{equation} \label{t2}
t=t_0\int_0^Wdx\,\psi_w^*(x)\psi_{s}(x).
\end{equation}
Quantization along the width gives the wave functions $\psi_{w,s}(x)=\sqrt{2/W}\sin(\pi n_{w,W}x/W)$, where $n_{w,W}\in\mathbb{Z}^+$ are the quantum numbers for transverse subbands in the wire and superconductor, respectively. Evaluating the integral in Eq.~\eqref{t2}, we find that only transverse subbands with the same quantum number couple to each other,
\begin{equation}
t=t_0\delta_{n_w,n_W}.
\end{equation}
To second order in tunneling, the induced gap on a given wire subband (characterized by $n_w$) takes the form
\begin{equation} \label{pt}
E_{g,n_w}(d,W)\propto\sum_{n_d,n_W}\frac{|t|^2}{E_{n_d,n_W}},
\end{equation}
where $n_d$ and $n_W$ are quantum numbers characterizing the spectrum of the superconductor (due to the finite thickness $d$ and finite width $W$, respectively), which is given by 
\begin{equation} \label{Esc}
E_{n_d,n_W}=\sqrt{\left(\mu_s-\frac{\hbar^2\pi^2n_d^2}{2m_sd^2}-\frac{\hbar^2\pi^2n_W^2}{2m_sW^2}\right)^2+\Delta^2}.
\end{equation}
In Eq.~\eqref{Esc} we neglect the momentum dependence of the spectrum, as we assume that only momenta $k\ll k_{Fs}$ are relevant.

As the quantum wire has at most only a few occupied subbands, this restricts $n_W\sim1$. Furthermore, since relevant $n_d\sim50$ (determined by requiring $\mu_s\sim\hbar^2\pi^2n_d^2/2m_sd^2$ and taking $\mu_s\sim10$ eV and $d\sim10$ nm) and $W\gg d$, we have $n_W^2/W^2\ll n_d^2/d^2$. Provided that $|\mu_s-\hbar^2\pi^2n_d^2/2m_sd^2|\gg\hbar^2\pi^2n_W^2/2m_sW^2$, which is true for almost all $d$, the term containing $W$ in Eq.~\eqref{Esc} is negligible. Performing the sum over $n_W$ then gives 
\begin{equation}
E_{g,n_w}(d,W)\propto\sum_{n_d}\frac{|t_0|^2}{E_{n_d}},
\end{equation}
where
\begin{equation}
E_{n_d}=\sqrt{\left(\mu_s-\frac{\hbar^2\pi^2n_d^2}{2m_sd^2}\right)^2+\Delta^2}
\end{equation}
is the spectrum of the superconductor in the limit $W=0$. We see that both $W$ and $n_w$ have dropped out of the expression for the gap completely. Hence, the induced gap is the same for all subbands of the wire and is independent of the width $W$.

\begin{figure}[b!]
\centering
\includegraphics[width=.8\linewidth]{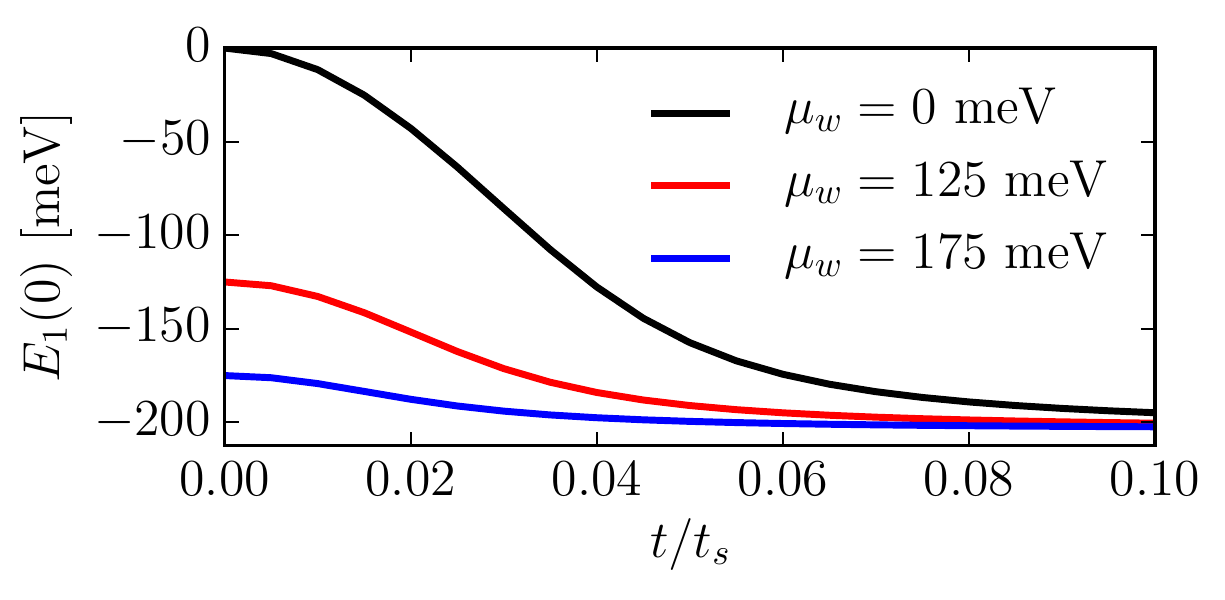}
\caption{\label{shiftvsmu} Energy of lowest subband at $k=0$, $E_1(0)$, plotted as a function of tunneling strength $t$ for various wire chemical potentials $\mu_w$. While the band shift $\delta E_1$ is dependent on $\mu_w$, the bottom of the band always approaches the same energy in the strong-coupling limit. This indicates that the spectrum is determined entirely by the superconductor in this limit. Tight-binding parameters are the same as in Fig.~\ref{deltamuvsd}.}
\end{figure}

\section{Additional calculations for epitaxial Al/InAs} \label{AppendixB}

In Sec.~\ref{sec2D}, we discussed how the proximity-induced gap and band shift behave as a function of superconductor thickness $d$ in the strong-coupling limit. Here, we demonstrate that  the parameters of the wire are significantly renormalized by the tunnel coupling. Our results are displayed in Fig.~\ref{real2D}. For realistic experimental parameters, we find even more drastic changes in semiconducting properties than in Fig.~\ref{specvst}. For $t=0.1t_s$, which was the tunneling strength used to generate Fig.~\ref{deltamuvsd}, the effective mass of the lowest subband is $m^*\sim0.3m_e$, the spin-orbit splitting is $E_{so}\sim10$ $\mu$eV, and the $g$-factor is $|g|\sim2$. [While the effective mass was previously deduced by fitting to Eq.~\eqref{spectrum}, here, we determine it  as  $m^*=\hbar^2k_{so}^2/2E_{so}$, where $k_{so}$ is the momentum at which the band attains its minimum.] As shown in Sec.~\ref{secStrongCoupling}, the higher subbands that lie closer to the chemical potential will have a slightly weaker parameter renormalization, which is why we quote slightly different values while making estimates in Sec.~\ref{secTopological}.

\section{Strong-coupling spectrum independent of wire chemical potential $\mu_w$} \label{AppendixC}

In all calculations of the main text, we have assumed that the chemical potential of the wire is tuned to the Rashba crossing point of the lowest transverse subband ($\mu_w=0$) at $t=0$. In reality, however, it is not known how many transverse subbands are occupied in the wire and it is possible that $\mu_w$ takes a different value. It is thus important to test whether our main result, namely the shift of the lowest transverse subband to large energies induced by the superconductor in the strong-coupling limit, is affected by our choice of $\mu_w$. 

The energy of the lowest subband at $k=0$, $E_1(0)$, is plotted as a function of $t$ for various $\mu_w$ (ranging between 0 and 175 meV) in Fig.~\ref{shiftvsmu} for the 2D (single subband) case. While the band shift $\delta E_1$ is dependent on $\mu_w$, $E_1(0)$ converges to the same energy regardless of the initial position of the wire chemical potential $\mu_w$.

\begin{figure}[t!]
\centering
\includegraphics[width=\linewidth]{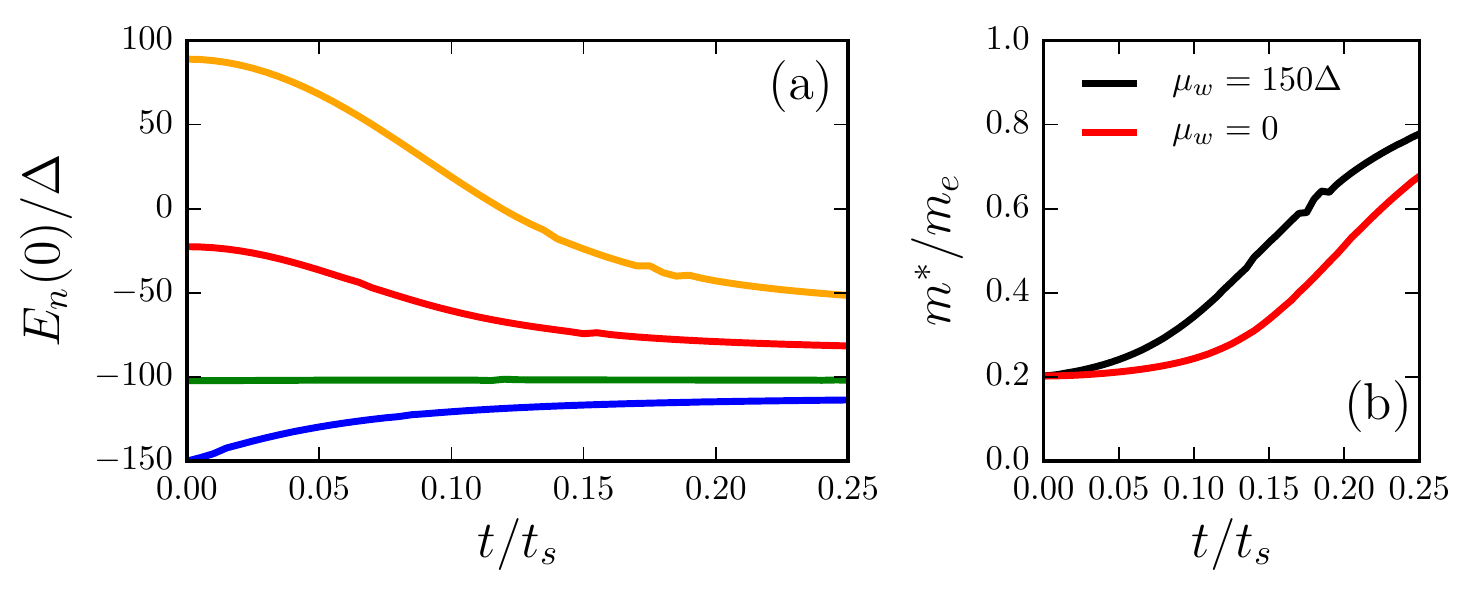}
\caption{\label{deltamumanybands} (a) Energy of four lowest transverse subbands at $k=0$, $E_n(0)$, plotted as a function of tunneling strength $t$. Parameters are the same as in Fig.~\ref{specweak}, with $\mu_w=150\Delta$. Comparing with Fig.~\ref{specvst}(b) (which has $\mu_w=0$), we see that all four subbands approach the same energy in the limit of large $t$ regardless of $\mu_w$. (b) Effective mass $m^*$ for both $\mu_w=150\Delta$ (black curve) and $\mu_w=0$ [red curve, corresponding to Fig.~\ref{specvst}(c)]. Because the band shifts are smaller when $\mu_w=150\Delta$, the mass is renormalized more quickly as a function of $t$.}
\end{figure}

Similarly, the energy of the four lowest transverse subbands at $k=0$, $E_n(0)$, is plotted in Fig.~\ref{deltamumanybands}(a) as a function of $t$ for the 3D case when there are several occupied subbands in the limit $t=0$. Comparing with Fig.~\ref{specvst}(b), we find that the energies of all four subbands approach the same values in the strong-coupling limit. However, because the band shifts are smaller for the case of several occupied subbands, the material parameters are more quickly renormalized as a function of $t$ [e.g., renormalization of the effective mass $m^*$ is shown in Fig.~\ref{deltamumanybands}(b)].

These results indicate that the spectrum of the system is determined entirely by the superconductor in the strong-coupling limit. Thus, regardless of how many subbands are occupied in the wire to begin with, the metallization picture presented in Fig.~\ref{renormalization}(b) still holds. 

\bibliography{bibStrongCouplingDisorder}

\begin{thebibliography}{137}
\bibitem [{str()}]{strongcoupling}%
  \BibitemOpen
  \href@noop {} {}\bibinfo {note} {More specifically, the strong-coupling limit
  (defined when the induced gap satisfies $E_g\sim\Delta$), is obtained for
  $\pi t^2\nu_s\gtrsim\Delta$, where $\pi t^2\nu_s$ is the usual tunneling
  energy scale ($t$ is a tunneling amplitude and $\nu_s$ is the density of
  states of the metal).}\BibitemShut {Stop}%
\bibitem [{ras()}]{rashba}%
  \BibitemOpen
  \href@noop {} {}\bibinfo {note} {For simplicity, we assume that the
  spin-orbit coupling is intrinsic to the semiconductor. However, any
  modifications to the spin-orbit strength by the interface with the
  superconductor can be accounted for by renormalizing $\alpha$.}\BibitemShut
  {Stop}%
\bibitem [{tb()}]{tb}%
  \BibitemOpen
  \href@noop {} {}\bibinfo {note} {We choose the hopping amplitude in the
  superconductor to be a factor of 10 larger than the chemical potential in
  order to model the continuum free-electron model, so that there are
  negligible band curvature effects impacting our numerical
  results.}\BibitemShut {Stop}%
\bibitem [{gfa()}]{gfactor}%
  \BibitemOpen
  \href@noop {} {}\bibinfo {note} {We have verified numerically that the
  renormalization of the $g$-factor does not depend on the specific choice for
  $\Delta_Z$ and is also not affected by superconductivity (\emph{i.e.}, it is
  the same for $\Delta=0$).}\BibitemShut {Stop}%
\bibitem [{fie()}]{fielddependence}%
  \BibitemOpen
  \href@noop {} {}\bibinfo {note} {We note that we do not treat
  superconductivity self-consistently in the presence of the magnetic field as
  done, for example, for magnetic impurities on superconducting surfaces
  \cite{Flatte_1997,Balatsky_2006,Meng_2015}. However, this can only decrease
  the order parameter in the superconductor, which would not affect the band
  shift or renormalization of the semiconductor parameters.}\BibitemShut
  {Stop}%
\end{thebibliography}

\end{document}